\documentclass[%
 reprint,
superscriptaddress,
 amsmath,amssymb,
 aps,
 pre,
]{revtex4-1}

\usepackage{graphicx}% Include figure files
\usepackage{dcolumn}% Align table columns on decimal point
\usepackage{bm}% bold math
\usepackage[colorlinks= true, linkcolor=blue, citecolor=blue, urlcolor=blue]{hyperref}% add hypertext capabilities
\usepackage{lipsum}
\usepackage{bbold}
\usepackage{mathtools}
\usepackage[normalem]{ulem}

\def\bea{\begin{eqnarray}}
\def\eea{\end{eqnarray}}

\usepackage{xcolor}

\begin{document}

%\preprint{APS/123-QED}

\title{Universal framework for record ages under restart}

\author{Aanjaneya Kumar}
\email{kumar.aanjaneya@students.iiserpune.ac.in}
\affiliation{Department of Physics, Indian Institute of Science Education and Research, Dr. Homi Bhabha Road, Pune 411008, India.}
\author{Arnab Pal}
\email{arnabpal@imsc.res.in}
\affiliation{The Institute of Mathematical Sciences, CIT Campus, Taramani, Chennai 600113, India}
\affiliation{Homi Bhabha National Institute, Training School Complex, Anushakti Nagar, Mumbai 400094, India}

%\date{\today}

\begin{abstract}
We propose a universal framework to compute record age statistics of a stochastic time-series that undergoes random restarts. The proposed framework makes minimal assumptions on the underlying process and is furthermore suited to treat generic restart protocols going beyond the Markovian setting. After benchmarking the framework for classical random walks on the $1$D lattice, we derive a universal criterion underpinning the impact of restart on the age of the $n$th record for generic time-series with nearest-neighbor transitions. Crucially, the criterion contains a penalty of order $n$, that puts strong constraints on restart expediting the creation of records, as compared to the simple first-passage completion. The applicability of our approach is further demonstrated on an aggregation-shattering process where we compute the typical growth rates of aggregate sizes. This unified framework paves the way to explore record statistics of time-series under restart in a wide range of complex systems.
\end{abstract}

\pacs{Valid PACS appear here}% PACS, the Physics and Astronomy
\begin{titlepage}
\maketitle
\end{titlepage}

\emph{Introduction.---}How long will it take for the price of a stock to cross its current all-time-high value? When will another human being cover a 100 metres faster than Usain Bolt? These questions pertain to computing record ages, a quantity that lies at the heart of the subject of record statistics \cite{nevzorov_records_1988,gulati2003parametric,wergen_records_2013,schehr_exact_2013,godreche_record_2017,sabhapandit_extremes_2019}. The study of record-breaking events has generated immense research interest since the pioneering work of Chandler in 1952 \cite{chandler_distribution_1952}, owing to its applications in fields including finance \cite{wergen_record_2011,sabir_record_2014,santhanam2017record}, climate studies \cite{hoyt_weather_1981,schmittmann_weather_1999,benestad_how_2003,redner_role_2006}, hydrology \cite{vogel_frequency_2001}, sports \cite{gembris_trends_2002,gembris2007evolution}, and also physics \cite{alessandro1990domain,sabhapandit2011record,majumdar_record_2012,godreche_universal_2014,godreche_record_2015,majumdar2019exactly}. 

The prototypical setting in the study of records consists of a discrete time-series $\bar{x} = \{x_0, x_1, x_2, \dots \}$, where the entries could represent the daily temperatures of a city, the number of people infected in a day during a pandemic, or any other observable of interest which is being measured at discrete time points. The $j$-th entry, $x_j$ of the time-series $\bar{x}$ is called a record if its numerical value exceeds the values of all preceding entries $x_i$, i.e., $x_j>x_i$ for all $i<j$. Important insight into the persistence of a record $x_i$ is obtained through the record age $L(x_i)$, which denotes the number of time-steps needed for a new record to be created, after $x_i$.  Concretely, for two consecutive records $x_{i}$ and $x_{j}$, the record age $L(x_i)$ is defined to be $j-i$, as depicted in Fig.~\ref{fig:schematic}(a), where the two yellow symbols denote record events.

\begin{figure}[t]
    \centering
    \includegraphics[width=0.9\columnwidth,height=6.2cm]{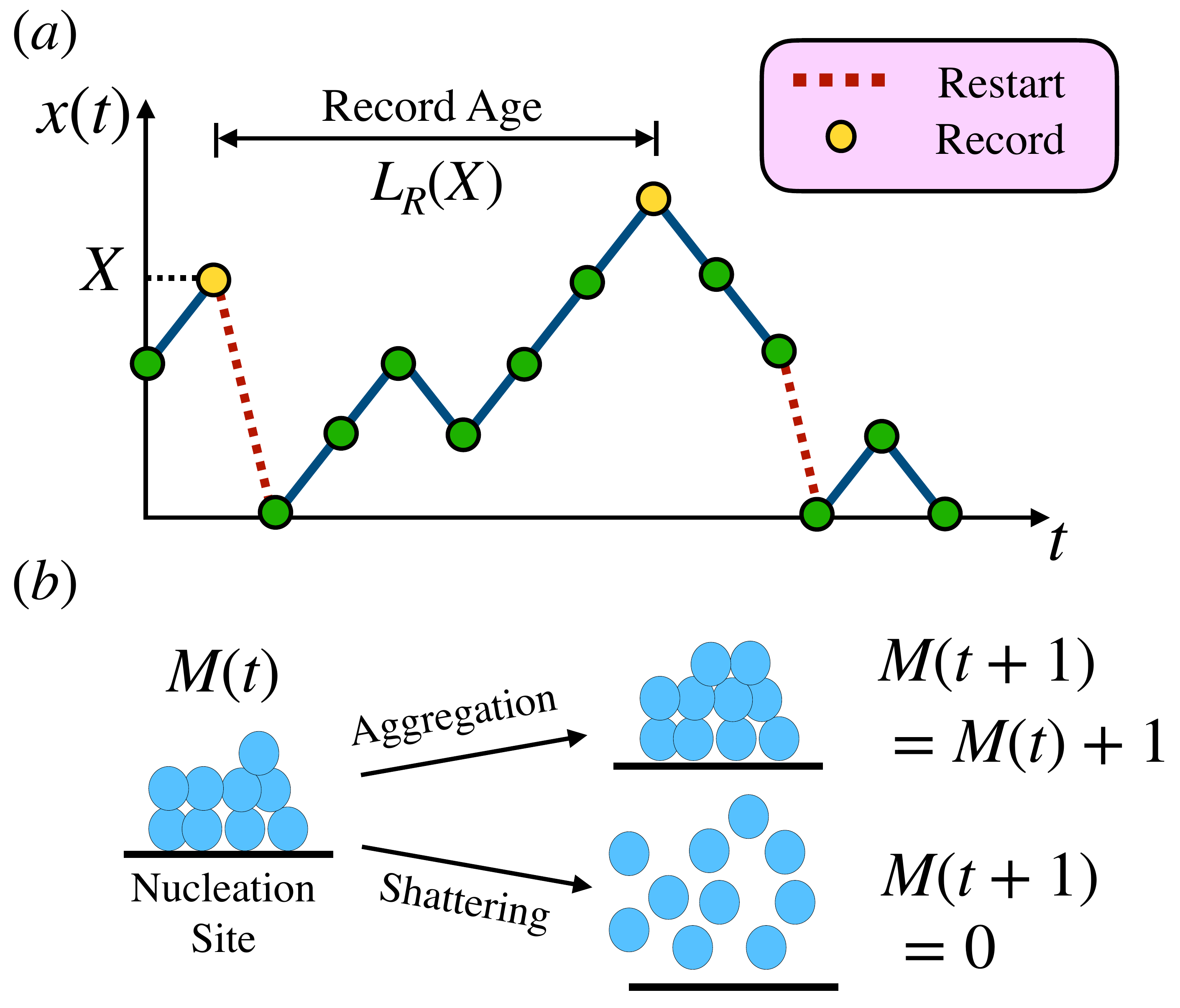}
    \caption{(a) Schematic for an observable $x(t)$ that evolves and undergoes stochastic resetting. Yellow points denote record events ($X$). Dashed red lines denote restart events, and blue lines denote non-restart transitions in the evolution of the process. (b) A mass aggregation-shattering process. Masses aggregate  on a nucleation site until a shattering event occurs, resetting the mass to zero. Aggregation resumes and continues to grow until the next shattering event.}
    \label{fig:schematic}
\end{figure}

While most of the efforts have been focused towards studying the case where the entries of the time-series $\bar{x}$ are independent and identically distributed (IID) random variables, oftentimes the entries obtained from real-world scenarios are in fact correlated. Moreover, as we live in a world where global catastrophes seem to be inevitable, consequently their signatures appear in most data sets of practical relevance, with examples including a sudden fall in the price of a stock \cite{stojkoski2022income}, sharp layoff of individual jobs due to post-pandemic recession, or a massive extinction of population due to catastrophe \cite{di_crescenzo_note_2008}. This ubiquitous feature of resetting events is not only limited to economics \cite{stojkoski2022income,gabaix2016dynamics}, operations research \cite{bonomo2021mitigating} or ecology \cite{pal_search_2020}, but can also be observed in microscopic out-of-equilibrium physical \cite{evans_diffusion_2011,pal2015diffusion,evans_stochastic_2020,gupta2020work,gupta2014fluctuating}, chemical \cite{reuveni2014role,robin2018single}, or biological \cite{roldan2016stochastic,budnar2019anillin} systems. More recently, restart has also emerged as an efficient  strategy to speed-up complex search processes with potential applications in optimization problems \cite{luby1993optimal,montanari2002optimizing,gomes1998boosting,huang2007effect} and search theory \cite{evans_diffusion_2011,kusmierz2014first,reuveni_optimal_2016,pal_first_2017,ray2019peclet,pal_first_2019,chechkin2018random,pal_search_2020,evans_stochastic_2020,gupta2022stochastic,singh_extremal_2021,de2020optimization,huang2021random,ray2021mitigating,riascos2020random,ye2022random,campos2015phase}. A natural question then arises: How do such restart events ramify the record statistics -- in particular, the record ages? Quite  remarkably, our answer to this question also sheds light on a seemingly unrelated problem namely the lifetime statistics in the mass aggregation-shattering models (see Fig.~\ref{fig:schematic}).

The central theme of this Letter is to build a unified formalism that allows us to obtain record ages for time-series generated by arbitrary stochastic processes which are subjected to intermittent collapse-restart events. Employing ideas and techniques from the first-passage under restart description \cite{pal_first_2017,bonomo2021first,bonomo2021p}, we distill the core principles that underpin the universal behavior of record ages under arbitrary restart. This allows us to probe record ages in a very generic setting covering both Markov and non-Markov processes, with minimal assumptions. In particular, we derive a universal criterion that dictates the effect of restart on the record ages.
Notably, the statistics of the number of records (average properties) have been studied recently for random walk (RW) models under the assumption of geometric restart steps \cite{majumdar2021record,godreche2022maximum}. However, the observable of interest herein is the record ages which have not been studied hitherto. After demonstrating the formalism for a biased RW, we apply it to the widely applicable aggregation-fragmentation models (Fig.~\ref{fig:schematic}(b)). To be specific, we compute the growth rate for mass-aggregates that requires us to generalize the formalism to arbitrary shattering/restart events that are not necessarily rate limiting process but can also have intrinsic temporal heterogeneity.

\emph{General formalism.---}We start by considering an extremely general case, where we have an arbitrary discrete time-series $\bar{x} = \{x_1, x_2, x_3 \dots \}$ generated by a stochastic process. Corresponding to this time-series, we have the set of records $\bar{X} = \{X_1, X_2, X_3, \dots \}$, where $X_i$ denotes the numerical value of the $i$-th record breaking event in the time-series $\bar{x}$. For each $X_i$, we define the record age $L(X_i)$ to be the time taken for the next record-breaking event $X_{i+1}$ to occur following $X_i$ \cite{note1}.

Now, suppose the stochastic process generating the time-series $\bar{x}$ is subjected to random restart events, whose occurrence bring the numerical value of the subsequent entry in the time-series to a predetermined value that is assumed to be $0$. Note, however, a generalization to this assumption (i.e., restart from another arbitrary value or from an ensemble) is feasible within our framework. 
Let us denote by $X$ an entry that is a record-breaking event in the time-series generated by the stochastic process under restart events. For simplicity, let us assume that these restart events take place after some geometrically distributed random time-step $R$ (generalization to arbitrary distributions is considered later). The age of the record $X$ (under restart) is denoted by  $L_R(X)$. If the record-breaking event subsequent to the formation of record $X$ occurs prior to any restart, we have $L_R(X)=L(X)$. Otherwise, the process resets to $0$ after time $R$, and from there the resultant process has to be observed until it crosses the record $X$. Combining these two possibilities, one has
\begin{equation}
    L_R(X) = \begin{dcases}
    L(X)& \text{if } L(X) < R\\
    R +  T^{R}_{X,0}   & \text{otherwise,}
    \end{dcases}
\label{renewal}
\end{equation}
where $T_{X,0}^{R}$ is the time taken for the time series to cross the threshold $X$ for the first time, given that it starts from $0$, in the presence of restart events. Equation~\eqref{renewal} is central to our analysis. Indeed, noting that $ L_R(X) = \text{min}\{L(X),R\} + \mathbb{1}\big(R\leq L(X)\big)T^{R}_{X,0}$, where $\text{min}\{z_1,z_2\}$ is the minimum of $z_1$ and $z_2$ and $\mathbb{1}\left( z_1 \leq z_2 \right)$ is an indicator random variable which is unity if $z_1 \leq z_2$ and zero otherwise, we find the mean record age as follows
\begin{equation}
    \langle L_R(X) \rangle = \langle \text{min}\{L(X),R\}  \rangle + \text{Pr}\big(R\leq L(X)\big) \langle T^{R}_{X,0} \rangle,
    \label{meanLn1}
\end{equation}
where $\langle T^{R}_{X,0} \rangle$ is the mean first-passage time under restart \cite{bonomo2021first,bonomo2021p,flynn2022using}.
Given the statistics of individual terms, one can then compute the mean record age using Eq. \eqref{meanLn1}. Notably, Eq.~\eqref{renewal} serves as a backbone to provide the full statistics of the record age which is also a perceived challenge. To gain further insights, we first illustrate our formalism on the 1D lattice RW and then show how the generalized theory applies to more complex scenarios. 
\begin{figure}
    \centering
    \includegraphics[height=3.4cm, width=8.5 cm]{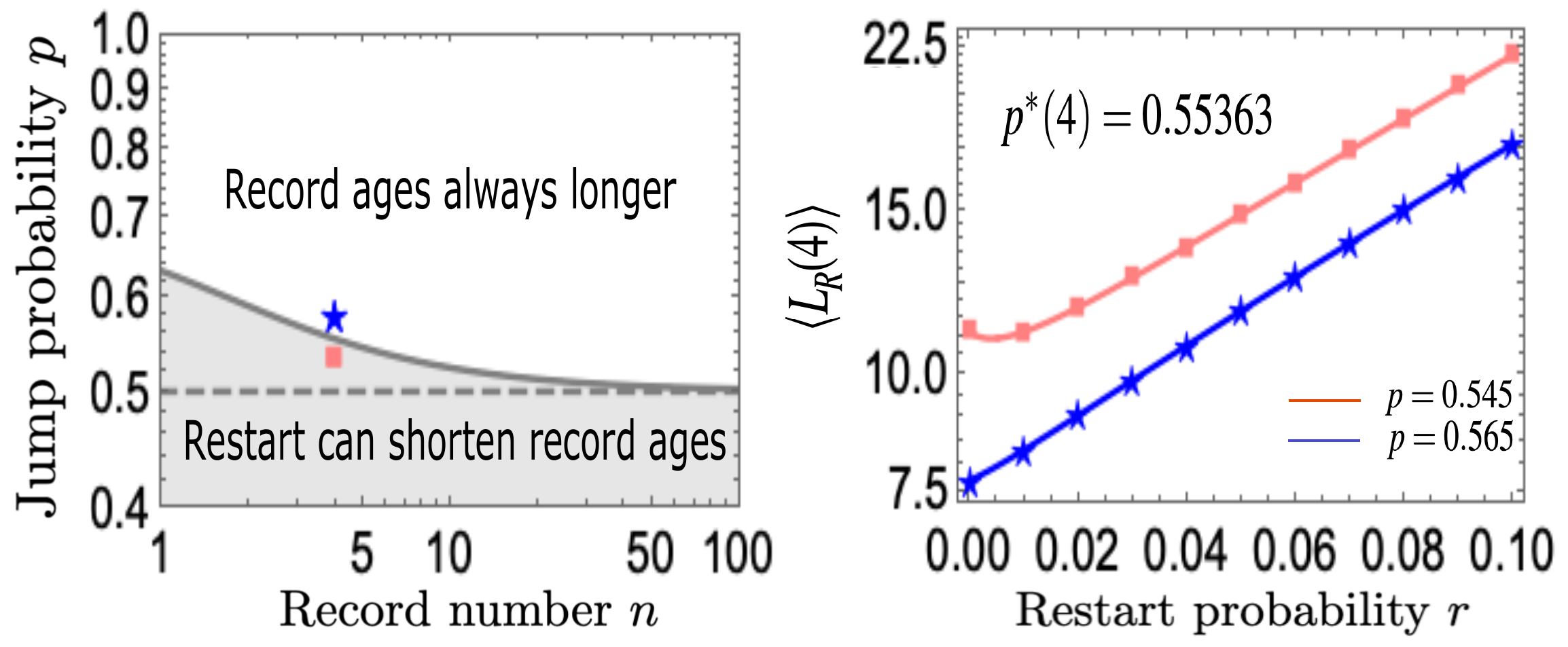}
    \caption{Left: Phase diagram 
    for record ages in a biased RW on a 1D lattice. Right: Mean age of the $4$th record $\langle L_R(4) \rangle$ as a function of restart probability $r$ for $p=0.565 (>p^*(4))$ (blue) and  $p=0.545 (<p^*(4))$ (orange), where $p^*(n) $ is the value of the bias $p$ (taken from the phase-separatrix on the left panel for a given $n$ [see Sec S2 in \cite{SI}]), beyond which restart cannot shorten record ages. Here, $p^*(4) = 0.55363$. The solid lines are obtained from our \textit{analytical} formula, while symbols represent values obtained from \textit{numerical simulations}. }
    \label{fig:fig2}
\end{figure}

\emph{Random walks on 1D lattice.---}A major advancement in our understanding of record statistics beyond IID random variables has come through the example of RW which were popularized following Pólya’s seminal work \cite{polya1921aufgabe}. A major advantage of using random walk models is that we can gain much insights by solving them analytically \cite{montroll1965random,klafter2011first,hughes1995random,giuggioli2020exact}.
To proceed further, we assume that the 1D RW evolves with the dynamics $x_{i}=x_{i-1} + \eta_{i-1}$ where $x_i$ denotes the position of the RW at $i$-th step and $\eta_i$ is the increment. The walker is biased so that $\eta_i=+1$ with probability $p$, and $\eta_i=-1$ with probability $1-p$, for all $i$. Positions of the RW ($x_i$) represent a \emph{strongly correlated} time-series. Furthermore, the walker experiences sharp transitions with probability $r$ to the origin after which it restarts its dynamics \cite{kusmierz2014first,bonomo2021first}.

For the case of random walk on a $1$D lattice with nearest neighbour jumps, the time-series of the position of the walker $\bar{x}$ is a sequence of integers, and consequentially the same holds for the sequence of records $\bar{X}$. We denote by $L_R(n)$ the time taken for a record-breaking event to occur, after the last record was created at position $n$. In the absence of resetting, the record age $L(n)$ is simply $T_{n+1,n}$ -- the first-passage time to go from $n$ to $n+1$, and it is independent of $n$. However restart introduces an inherent heterogeneity in the problem so that the record ages depend on the record number, or the magnitude of the last record. 
To see this, we first obtain the mean from Eq.~\eqref{meanLn1} 
\begin{equation}
 \langle L_R(n) \rangle = \langle \text{min}\{T_{n+1,n},R\} \rangle + \text{Pr}\big[ R\leq T_{n+1,n} \big]  \langle T^{R}_{n+1,0} \rangle,   
\end{equation}
where each component of the RHS can be computed given the distribution of $T$ and $R$. In particular, for geometric distribution of restart steps, we have \cite{SI}
\begin{equation}
     \langle L_R(n) \rangle = \frac{\widetilde{Q}_{1-r}(n+1|n)}{1-r\widetilde{Q}_{1-r}(n+1|0)},
     \label{mean-RW-r}
\end{equation}
where $\widetilde{Q}_{z}(i|j) = \sum_{k=0}^{\infty}z^k Q_k(i|j)$ is the generating function of the survival probability
$Q_k(i|j)$ that denotes the probability for a RW starting from a site $j<i$ and \emph{not} reaching $i$ till the $k$-th time-step. It is important to note that $\langle L_R(n) \rangle$ is expressed solely in terms of the survival properties for the bare process. Furthermore, the survival probability equals $\tilde{Q}_z(i|j)=\frac{1-\tilde{F}_{z}(i|j)}{1-z}$
where $\tilde{F}_{z}(i|j)$ denotes the generating function of the first-passage time distribution $F_{k}(i|j)=\text{Prob}[T_{i,j}=k]$ that the walker starts from state $j$ and reaches $i$ for the first time exactly in $k$ steps. For the biased RW, this can be expressed as \cite{klafter2011first}
\begin{align}
        \tilde{F}_{z}(i|j) =\left(\frac{1-\sqrt{1-4p(1-p)z^2}}{2(1-p)z}\right)^{i-j}
\end{align}
for $i-j>0$. Replacing the expressions in \eqref{mean-RW-r}, one finds the mean record age for the RW. 

\begin{figure}[b]
    \includegraphics[scale=0.29]{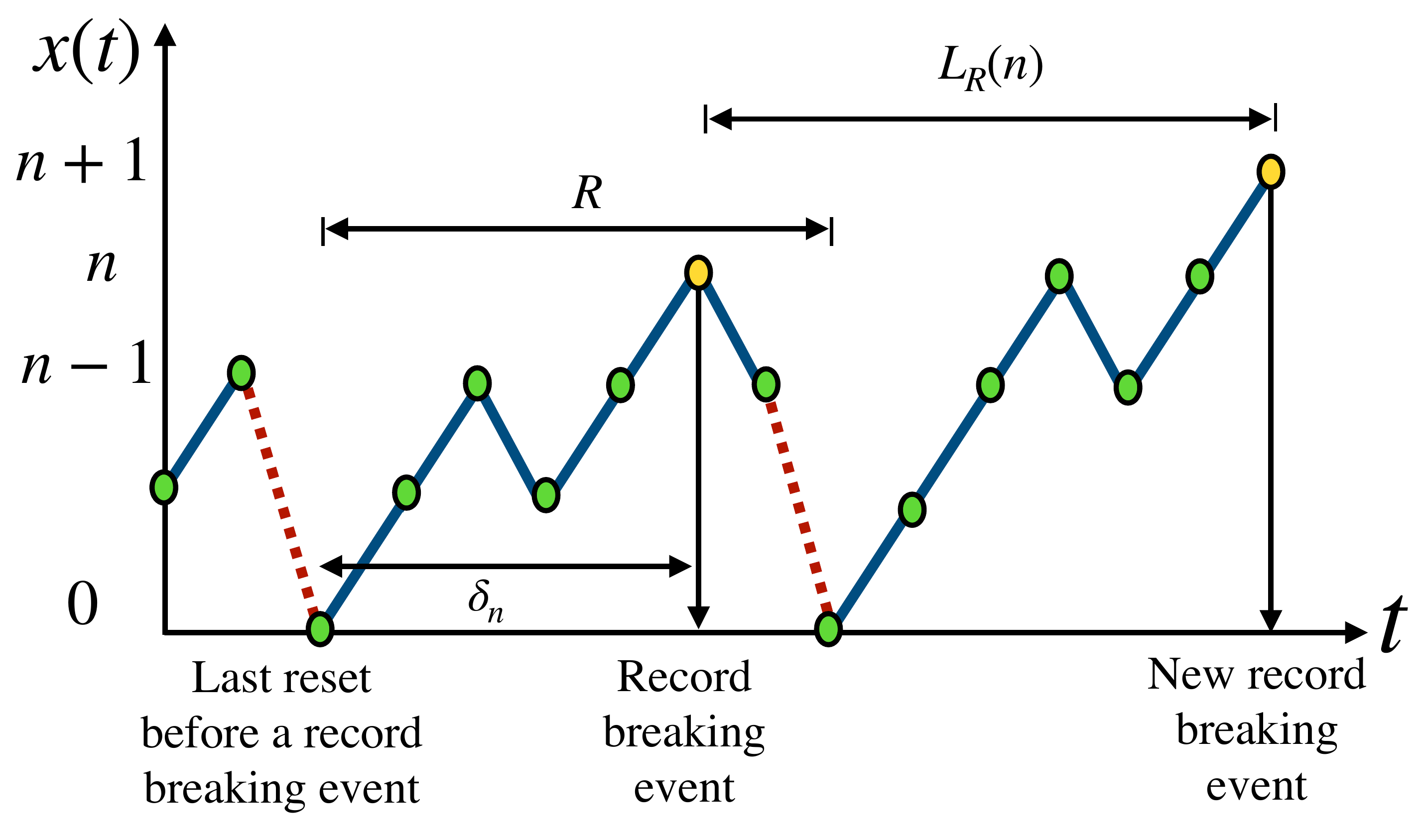}
    \caption{Schematic for record age statistics under non-geometric restarts. Unlike the Markovian case, the time taken for a restart event to occur after the creation of a record (in this figure, a record is created when the time-series reaches $n$ for the first time) is not simply given by the distribution of $R$, and instead is given by $R_{\delta_n}=R-\delta_n$.}
    \label{fig:general_renewal}
\end{figure}

Several comments are in order now. In the case of symmetric walkers for the bare process ($p=1/2,r=0$), the mean record age is infinite, and any restart probability $r<1$ would render the mean finite. In other words, collapse-restart events will always expedite record breaking events for symmetric RW. However, this need not be the case in general as restart could also result in longer record ages. Consider, for example, the biased RW, where the mean record ages are finite for the bare process. Thus, it is essential to pinpoint the transition point which can be understood by the introduction of an infinitesimal resetting probability. Indeed, expanding mean record age Eq.~\eqref{mean-RW-r} with respect to $\delta r \to 0$, one finds
$\langle L_{\delta r}(n) \rangle= \langle L(n) \rangle-\frac{\delta r}{2} \langle T_{n+1,n} \rangle^2 \left[ CV_{n+1,n}^2+\frac{\langle T_{n+1,n} \rangle-1-2\langle T_{n+1,0} \rangle}{\langle T_{n+1,n} \rangle}  \right]$, where $CV^2_{n+1,n} = \frac{\text{Var}(T_{n+1,n})}{\langle T_{n+1,n} \rangle^2}$
is the squared coefficient of variation of $T_{n+1,n}$. For restart to reduce the record age, one should have $\langle L_{\delta r}(n) \rangle  < \langle L(n) \rangle$ resulting in \cite{SI}
 \begin{align}
    CV^2_{n+1,n}&> \frac{2 \langle T_{n+1,0} \rangle +1}{\langle T_{n+1,n} \rangle}-1. \label{criterion}
\end{align}
Eq. \eqref{criterion} remarkably holds for any underlying stochastic process, and sets up a universal criterion for the effect of restart on record ages. In the paradigmatic case of biased RW, the criterion in Eq.~\eqref{criterion} reduces to 
 \begin{align}
    CV^2_{n+1,n}& > 2n + \left\{1+ \frac{1}{\langle T_{n+1,n}\rangle} \right\}, \label{criterion_biasedrw}
\end{align}
where the second term on the RHS is the criterion for the mean first-passage solely \cite{bonomo2021p}. Thus, the additional term of $2n$ corresponds to a ``penalty" for resetting to a point further away from the target, compared to the initial condition, setting up a very strict criterion on the relative fluctuations of the underlying first-passage process in order for restart to expedite record-breaking events. For large $n$, the criterion is dominated by the penalty term $2n$, as both $\langle T_{n+1,n} \rangle$ and $CV^2_{n+1,n}$ are independent of $n$, resulting in an invalid inequality. Thus, restart never shortens the record ages for large $n$. Based on Eq.~\eqref{criterion_biasedrw}, in Fig.~\ref{fig:fig2}(a) we illustrate the particular phase space region spanned by $p$ and $n$  where restart can expedite the creation of records (grey shaded). Note that for values of $p$ below the dashed line ($p=0.5$), restart renders $\langle L_R(n) \rangle$ finite for all $n$, and thus \emph{always} leads to shorter record ages. In panel (b), we further plot $\langle L_R(n) \rangle$ for $n=4$, as a function of restart probability $r$ for two different values of $p$:  (i) $p=0.565$ which chosen above the critical value $p^*(4) \approx 0.55363$ beyond which Eq.~\eqref{criterion_biasedrw} is not satisfied for $n=4$ \cite{SI}, and (ii) $p=0.545$ which lies below the critical value $p^*(4)$, demonstrating the validity of the criterion.

\textit{Record ages under arbitrary restart step.---}So far, we had restricted our discussion to geometric restarts. However, while going beyond this Markovian case is an important step, as evident through the first-passage literature \cite{pal2016diffusion,pal_first_2017,shkilev_continuous-time_2017, bodrova_continuous-time_2020}, it is also quite challenging. The key issue here is to know the statistics of the time required for a restart event to occur right after a record. While for Markovian set-up, this time coincides with the restart time ($R$) itself (due to the memoryless property of geometric restart events), it is generically different for arbitrary restart steps (see Fig.~\ref{fig:general_renewal} for a timeline illustration).

For such a stochastic process, we can identify a renewal structure for the record ages as the following 
\begin{equation}
    L_R(n) = \begin{dcases}
    L(n),& \text{if } L(n) < R_{\delta_n}\\
    R_{\delta_n} +  T^{R}_{n+1,0}   & \text{otherwise}
    \end{dcases}
\label{arb_reset}
\end{equation} 
where $R_{\delta_n} = \{R - \delta_n | R>\delta_n\}$ is the forward recurrence time, and $\delta_n$ is the backward recurrence time so that $\delta_n = \{ T_{n,0} |T_{n,0} <R\}
$ and $R_{\delta_n}+\delta_n=R$ [see Fig. (\ref{fig:general_renewal})]. The latter is distributed according to
\begin{align}
    P_{\delta_n}(k)
=F_k(n|0)\frac{\sum_{m=k+1}^{\infty}P_R(m)}{\text{Pr}\left(T_{n,0} < R\right)},\label{dx}
\end{align}
where $P_R(m)$ is the restart time density (not necessarily geometric).
We stress that while Eq.~\eqref{arb_reset} is written in terms of the variable $n$, keeping in mind discrete-state stochastic processes (e.g., RW), 
generalization to continuous state processes is straightforward.

For geometrically distributed restarts, $R_{\delta_n}$ and $R$ have statistically identical distribution and hence one recovers Eq.~\eqref{renewal}. \textcolor{black}{However, generically, $R_{\delta_n}$ pertains to a different distribution} \cite{SI}
\begin{equation}
    P_{R_{\delta_n}}(k) = \frac{\text{Pr}\left(R -  T_{n,0} = k  \right)}{\text{Pr}\left(R > T_{n,0}  \right) \cdot \text{Pr}(R>\delta_n)},
    \label{rdx}
\end{equation}
where $k$ takes strictly positive values. 
\begin{figure}
    \centering
    \includegraphics[scale=0.53]{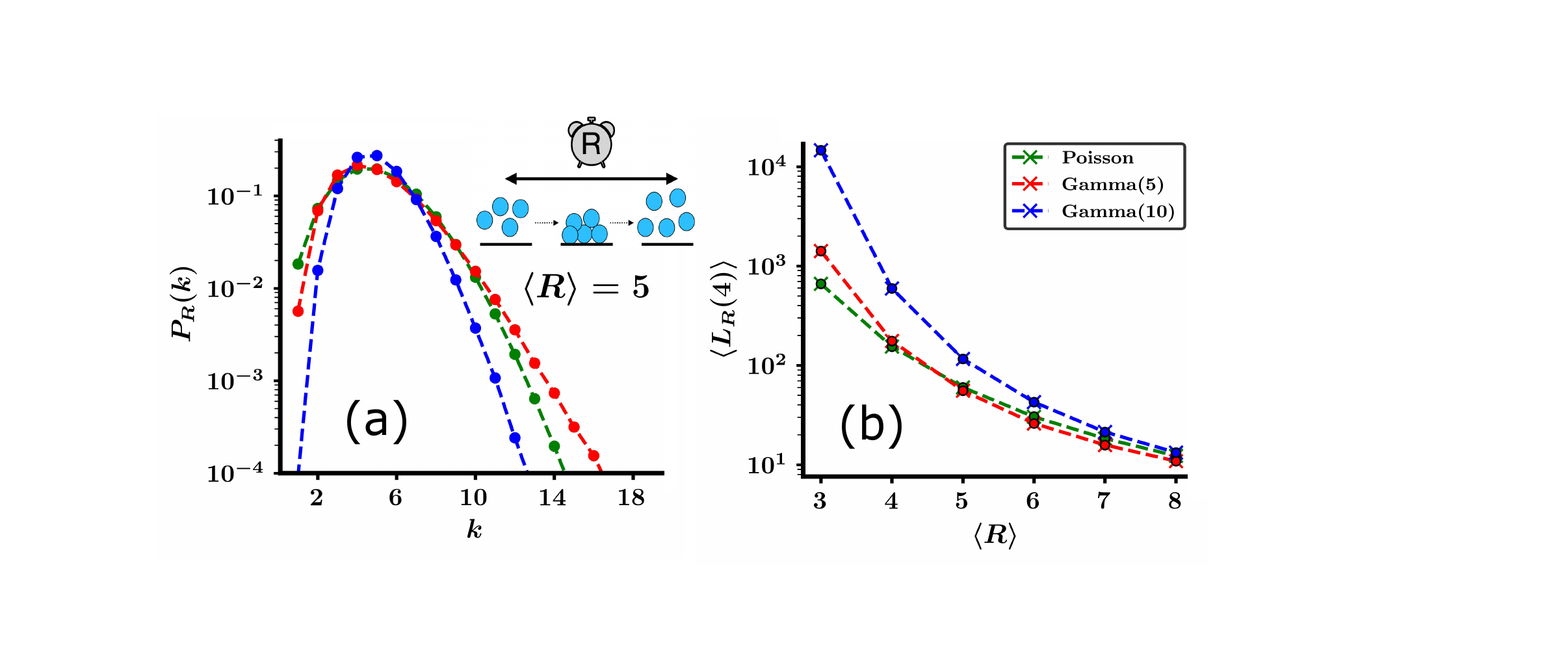}
    \caption{Mean record age in aggregation-shattering process. (a) Different non-Markovian restart distributions characterizing the time between shattering events. (b) Variation of $\langle L_R(4) \rangle$, under the restart distributions plotted in (a), as a function of mean restart time. The cross symbols denote our \textit{numerical} results and circles denote the \textit{theoretical} estimates from Eq.~\eqref{arb_meanLn1}.}
    \label{fig:fig4}
\end{figure}
Together, Eqs.~\eqref{arb_reset}, \eqref{dx}, and \eqref{rdx} allow us to write a closed set of equations to obtain the record age statistics of a time-series generated by an arbitrary stochastic process that undergoes possibly non-geometric restarts. For instance, the mean record age reads 
\begin{equation}
    \langle L_R(n) \rangle = \langle \text{min}\{L(n),R_{\delta_n}\}  \rangle + \text{Pr}\big(R_{\delta_n}\leq L(n)\big) \langle T^{R}_{n+1,0} \rangle,
    \label{arb_meanLn1}
\end{equation}
where we show that the mean record age under a non-geometric restart protocol can be expressed completely in terms of quantities related to the underlying process. This property holds also for all the subsequent moments of $L_R(n)$.

\emph{Aggregation-shattering processes.---} An important application of studying non-geometric (non-Markovian) restarts arises in the study of aggregation-fragmentation-shattering processes. Apart from being a paradigmatic model to probe non-equilibrium behaviour \cite{krapivsky_kinetic_2010,krapivsky_phase_2017,matveev_oscillations_2017,bodrova_kinetic_2019,brilliantov_nonextensive_2021}, models of aggregation and fragmentation have found diverse applications, ranging from modeling socio-economic phenomena \cite{ispolatov1998wealth,robson2021combined}  and neurodegenerative diseases \cite{thompson2021role,wang2009tau,fornari2020spatially}, to explaining the particle size distribution in saturn rings \cite{brilliantov_size_2015}, the distribution of sizes of animal groups in nature \cite{niwa_power-law_2003,ma_first_2011,nair_fission-fusion_2019} and raindrops \cite{srivastava1982simple}. In particular, in the case of neurodegenerative diseases, where it is argued that diseases like Alzheimer’s or Parkinson’s disease are caused by the pathological aggregation of certain proteins, it is suggested that some clearance mechanisms must also be at play, which keep these proteins from forming large aggregates in healthy individuals. These clearance mechanisms play the role of ``shattering" which bring down the size of aggregates \cite{thompson2021role}. Figure~\ref{fig:schematic}(b) is a schematic of such an aggregation-shattering process, where masses arrive, possibly non-geometrically, on a nucleation site, and form a larger aggregate. However, shattering of the aggregate can reset the mass at the nucleation site to zero.

Let us consider a time-series $M(t)$, which tracks the size of the aggregate at the nucleation site. Clearly, $M(t)$ is a stochastic process that undergoes restarts at random times. To delve deeper, let us 
assume that the inter-arrival times between two masses/monomers follow a geometric  distribution. Upon the arrival of a monomer, it sticks to the cluster of masses at the nucleation site (an aggregation event), leading to an increase in the mass at the site by unit one. However, `clearance' occurs at random times following possibly non-geometric distributions indicating a restart protocol with temporal memory. Clearance leads to the shattering of the cluster at the nucleation site, rendering $M(t)=0$ in that time-step. In this context, record age statistics of the aggregate size carries valuable insight into the lifetime of these masses, and the rate at which they grow.

In Fig.~\ref{fig:fig4} we demonstrate the dependence of $\langle L_R(4) \rangle$ i.e., the time until the formation of a mass-aggregate with $n=5$ units after an aggregate of size $4$ has been created for the first time, as a function of the mean shattering time $\langle R \rangle$, for non-geometric shattering times (drawn from Poisson distribution and discretized Gamma distribution \cite{SI}. The plot shows an excellent agreement between our theoretical prediction from Eq. (\ref{arb_meanLn1}) and the simulations. It is evident that shattering (restart) events slow down the process of creation of records, which concurs with our physical intuition. In \cite{SI}, we show how this is consistent with the criterion derived in Eq.~\eqref{criterion} and discuss other contrasting scenarios where such shattering-like mechanisms can expedite the creation of records despite the underlying time-series taking only non-negative values. Furthermore, note that for higher mass index (i.e., increasing $n$), the record age $\langle L_R(n) \rangle$ also keeps increasing which is a highlighted feature as well.

\emph{Conclusions.---}Record statistics has been a long standing focal point of research due to its numerous interdisciplinary applications that go beyond physics. In this work, we focus on understanding record statistics of a time-series that may contain signatures of catastrophes or sharp intermittent changes in the observed values. Modelling these events as restart, we build a unified framework to estimate record age statistics in a generic scenario. As such, our framework advances in encompassing arbitrary stochastic processes that undergo general non-Markovian restart events. Quite importantly, our framework reveals a universal criterion \eqref{criterion} that can predict the conditions under which restart could shorten the mean record ages of stochastic time-series with nearest-neighbor transitions. The application of this criterion is demonstrated not only for RW models where the underlying variable can be both positive and negative, but also for the mass aggregation model where the random variable remains strictly non-negative (see \cite{SI} for additional discussion).

Our work brings forward new insights on the intricate interplay between the inherent stochasticity pertaining to the system and the restart events. Although the focus has been on the average quantities, such ideas can also be extended to study fluctuations and higher moments. Finally, we highlight that sharp catastrophe in time-series \cite{stojkoski2022income} is a key signature of extreme events \cite{majumdar2020extreme,kishore_extreme_2011,malik_rare_2020} across complex systems. Thus, our formalism paves the way for building an improved understanding of rare events in natural systems and their consequences.

\emph{Acknowledgements.---}The authors gratefully acknowledge M. S. Santhanam for fruitful discussions. AK acknowledges the Prime Minister’s Research Fellowship of the Government
of India for financial support. AP gratefully acknowledges research support from the Department of Science and Technology, India, SERB Start-up Research Grant Number SRG/2022/000080 and Department of Atomic Energy, India.

\providecommand{\noopsort}[1]{}\providecommand{\singleletter}[1]{#1}%

\begin{titlepage}
\title{Supplemental material for ``Universal framework for record ages under restart"}
\maketitle
\end{titlepage}

\onecolumngrid
\setcounter{page}{1}
\renewcommand{\thepage}{S\arabic{page}}
\setcounter{equation}{0}
\renewcommand{\theequation}{S\arabic{equation}}
\setcounter{figure}{0}
\renewcommand{\thefigure}{S\arabic{figure}}
\setcounter{section}{0}
\renewcommand{\thesection}{S\arabic{section}}
\setcounter{table}{0}
\renewcommand{\thetable}{S\arabic{table}}

This Supplemental Material (SM) provides additional discussions and detailed mathematical derivations for the results mentioned in the main text. The SM is organized as follows: in Section~\ref{S0}, we set up the notation used throughout the manuscript as well as the SM, and define the key quantities of interest for quick reference. In Section~\ref{S1}, we provide a complete derivation of the record age statistics for the 1D random walk under geometric restarts as discussed in the main text. Following that, in Section~\ref{S2}, we derive a universal criterion for restart to expedite the creation of the $n$th record. The extension of our renewal framework to incorporate arbitrary restart mechanisms (beyond Markovian restart such as geometric) is presented in Section~\ref{S3}. We demonstrate the applicability of these general results in an aggregation-shattering model -- the details of which are discussed in Section~\ref{S4}.

% \tableofcontents

\section{Notation and definitions}\label{S0}

The typical setting considered in the manuscript is of a discrete time-series $\bar{x} = \{x_0, x_1, x_2, \dots \}$. The $j$-th entry $x_j$ of the time-series $\bar{x}$, is called a record if its numerical value exceeds the values of all preceding entries $x_i$, i.e., $x_j>x_i$ for all $i<j$. We denote by $\bar{X}$ the set of records $\bar{X} = \{X_1, X_2, X_3, \dots \}$, where $X_i$ denotes the numerical value of the $i$-th record breaking event in the time-series $\bar{x}$.

\begin{itemize}
    \item $L(X_i)$: \textcolor{black}{The record age $L(X_i)$, of the record with numerical value $X_i$ is defined to be the time taken for the next record-breaking event to occur after $X_i$}. For example, if record $X_i$ is created at time-step $t_1$ and the next record $X_{i+1}$ is created at time-step $t_2$, then $L(X_i) = t_2 - t_1$.
    \begin{itemize}
        \item For time-series generated by a lattice random walk with nearest-neighbour transitions, we will often denote the records using the standard notation $n$ for integers instead of $X$. Thus, $L(n)$ would denote the time taken for the time-series to reach the value $n+1$ after it has reached $n$ for the first time.
        \item \textcolor{black}{Interestingly, the above definition of $L(n)$ also coincides with another definition of record age, $\ell_n$, which is defined the time taken for the $(n+1)$-th record to be formed after the creation of the $n$-th record, given that the time-series starts from $0$ (this does not depend on the exact magnitude of the record). This is due to the fact that in a nearest-neighbour lattice random walk which starts from $0$, the $n$-th record happens when the time-series reaches value $n$.} 
    \end{itemize}
    \item  $L_R(X_i)$: The record age of the record $X_i$ under restart ($L_R(X_i)$) is the time taken for the next record-breaking event $X_{i+1}$ to occur after $X_i$, given that the time-series is subject to stochastic restart, whose occurrence resets the value of the time-series to a pre-defined quantity (say $0$). Evidently, $L_R(X_i)=L(X_i)$ in the absence of restart.
     \begin{itemize}
     \item In the case of lattice random walk with nearest-neighbour transitions, we denote the record ages under restart by $L_R(n)$, for the same reasons as mentioned above. 
     \end{itemize}
\end{itemize}

\section{Record age statistics for random walks under restart} \label{S1}
\subsection{General formalism}
Let us consider a random walk on a 1D lattice, with nearest neighbour jumps (i.e., $\eta_i=\pm 1$ for all $n$), subjected to geometric restarts to the origin. We will assign the probability for hopping later. We define $L_R(n)$ to be the time taken for the $(n+1)$-th record to be formed, after the $n$-th record was formed. Clearly, $L_R(n)$ is a random variable. For random walk without resetting, the record age $L(n)$ is independent of $n$, but as soon as resetting is introduced, there is an inherent heterogeneity in the problem, and the record ages start to depend on the record number, or the value of the last record. It can be seen that the random variable $L_R(n)$ satisfies the following renewal structure
\begin{equation}
    L_R(n) = \begin{dcases}
    L(n)& \text{if } L(n) < R\\
    R +  T^{R}_{n+1,0}  & \text{otherwise}
    \end{dcases}\label{SI_renewal}
\end{equation}
where $T_{i,j}^{R}$ and $T_{i,j}$ are random variables which denote the first passage time for the random walk from site $j$ to $i$ with and without resetting, and $R$ denotes the random time-step at which a resetting event happens and restates the walker to its initial coordinate. It is easy to see that Eq.\eqref{SI_renewal} can be expressed as
\begin{equation}
    L_R(n) = \text{min}\{T_{n+1,n},R\} + \mathbb{1}\big(R\leq T_{n+1,n}\big)T^{R}_{n+1,0}.
\end{equation}
where note that $L(n)=T_{n+1,n}$ is simply the record age for the underlying process and $\mathbb{1}\big(R\leq T_{n+1,n}\big)$ equals $1$ when $R\leq T_{n+1,n}$, and $0$ otherwise. Taking expectation on the both sides of the above equation yields
\begin{equation}
    \langle L_R(n) \rangle = \langle \text{min}\{T_{n+1,n},R\} \rangle + \langle \mathbb{1}\big(R\leq T_{n+1,n}\big)\rangle \langle T^{R}_{n+1,0} \rangle
    \label{SI_meanLn1}
\end{equation}
where the expectation value of the indicator function reads
\begin{equation}
    \langle \mathbb{1}\big(R\leq T_{n+1,n}\big) \rangle = \text{Pr}\big[ R\leq T_{n+1,n} \big].
\end{equation}
This allows us to write Eq.~\eqref{SI_meanLn1} as
\begin{equation}
    \langle L_R(n) \rangle = \langle \text{min}\{T_{n+1,n},R\} \rangle + \text{Pr}\big[ R\leq T_{n+1,n} \big]  \langle T^{R}_{n+1,0} \rangle,
    \label{SI_meanLn2}
\end{equation}
where notice that $\langle T^{R}_{n+1,0} \rangle$ is the simple mean first passage time (discrete) under restart (discrete) and can be computed using the framework of \emph{first-passage under restart}. Following \cite{1,2,3}, one has
\begin{equation}
   \langle T^{R}_{n+1,0} \rangle = \frac{\langle \text{min}\{ T_{n+1,0},R\}\rangle}{\text{Pr}\big[ T_{n+1,0} < R \big]},
   \label{FPUR-dis}
\end{equation}
which can be derived by noting that \cite{1}
\begin{equation}
\langle T^R_{n+1,0}\rangle = \langle \min\{T_{n+1,0},R\}\rangle + \langle \mathbb{1}(R\leq T_{n+1,0})\rangle \langle T^R_{n+1,0}\rangle.
\end{equation}
Plugging everything together in Eq.~(\ref{SI_meanLn2}), we find
\begin{equation}
    \langle L_R(n) \rangle = \langle \text{min}\{T_{n+1,n},R\} \rangle +  \frac{\text{Pr}\big[ R\leq T_{n+1,n} \big]}{\text{Pr}\big[ T_{n+1,0} < R \big]} \langle \text{min}\{T_{n+1,0},R\}\rangle,
    \label{SI_meanLn3}
\end{equation}
which is the mean record age for the random walk under restart. It is worth emphasizing that this result does not depend on the particular structure of random walk or the form of the resetting time density. 

\subsection{Geometric restart} 
In this section, we consider a specific form of restart time distribution, namely the geometric distribution. Here, a resetting step number is taken from the following distribution with parameter $r~  (0<r<1)$, 
\begin{align}
P_R(k)=(1-r)^{k-1} r, ~~k\geq 1.
\label{SI_geometric}
\end{align}
In other words, restart will occur exactly at the $k$-th step with probability $r$, after $k-1$ unsuccessful attempts. 
Notably, this distribution is the discrete analog of the exponential distribution, being a discrete distribution possessing the memory-less property.  To compute the mean record age for the random walk, let us now now evaluate Eq.~(\ref{SI_meanLn3}) term by term. \\

\noindent
\textbf{Evaluation of $ \langle \text{min}\{T_{n+1,n},R\} \rangle$:} We express $\langle \text{min}\{T_{n+1,n},R\} \rangle$ as \cite{2,3}
\begin{align}
    \langle \text{min}\{T_{n+1,n},R\} \rangle  &= \sum_{k=0}^{\infty} Q_k(n+1|n)(1-r)^k \nonumber \\
    &= \widetilde{Q}_{1-r}(n+1|n)
    \label{SI_min}
\end{align}
where in the first line, $Q_k(i|j)$ denotes the survival probability that the walker, starting from $j$, has not reached the site $i$ till the $k^\text{th}$ time-step. Furthermore in the second line, we have made use of its corresponding generating function defined as
\begin{align}
 \widetilde{Q}_{z}(n+1|n)  \equiv \sum_{k=0}^{\infty}z^{k}Q_{k}(n+1|n),
\end{align}
which has been used in the last line of Eq. (\ref{SI_min}) with the transformation $z \to 1-r$. Using a similar line of reasoning, we find 
\begin{align}
    \langle \text{min}\{T_{n+1,0},R\} \rangle  &= \sum_{k=0}^{\infty} Q_k(n+1|0)(1-r)^k\\
    &= \widetilde{Q}_{1-r}(n+1|0).
\end{align}
\\

\noindent
\textbf{Relation between the survival function and first passage time density:} Before proceeding further, it will be useful to recall the relation between the survival function and the first passage time density. By definition, they are connected to each other via
\begin{align}
%  Q_{j-1}(n+1|0)&=\sum_{k = j}^{\infty} F_k(n+1|0)   \\
 Q_k(i|j)&=\text{Pr}\left[ T_{i,j}>k \right]=\sum_{l = k+1}^{\infty} F_l(i|j),
\label{fqrelation-2}
\end{align}
which translates to the following in the $z$-space \cite{3,4}
\begin{equation}
    % \widetilde{F}_{1-r}(n+1|0)=1-r\widetilde{Q}_{1-r}(n+1|0), \\
    % &G_{Q_{N}}(z)&\equiv \sum_{n=0}^{\infty}z^{n}Q_{N}(n)=\frac{1-G_{N}(z)}{1-z},\\
    \widetilde{{Q}_{z}}(i|j)\equiv \sum_{k=0}^{\infty}z^{k}Q_{k}(i|j)=\frac{1-\widetilde{F}_{z}(i|j)}{1-z},
    \label{fqrelation}
\end{equation}
where $\tilde{F}_z(i|j)=\sum_{k=0}^{\infty}z^{k}F_{k}(i|j)$ is the probability generating function for first passage time density given by $F_{k}(i|j)=\text{Pr}[T_{i,j}=k]$.  Unlike survival function, first passage time density $F_{k}(i|j)$ computes the time-step $k$ at which the process reaches the site $i$ for the first time starting from $j$. This relation will be extensively used in below.
\\

\noindent
\textbf{Evaluation of $\text{Pr}\big[ T_{n+1,0} < R \big]$:}
Using Eq. (\ref{fqrelation}), we can now compute the term $\text{Pr}\big[ T_{n+1,0} < R \big]$ as follows
\begin{align}
    \text{Pr}\big[ T_{n+1,0} < R \big] &= \sum_{j=0}^{\infty} F_j(n+1|0)\sum_{k = j+1}^{\infty} (1-r)^{k-1}r, \\
     &= \sum_{j=0}^{\infty} F_j(n+1|0) \cdot (1-r)^j\\
    &= \widetilde{F}_{1-r}(n+1|0)=1-r\widetilde{Q}_{1-r}(n+1|0).
\end{align}
\\

\noindent
\textbf{Evaluation of $  \text{Pr}\big[ R\leq T_{n+1,n} \big] $:}
Following the above steps, we compute
\begin{equation}
    \text{Pr}\big[ R\leq T_{n+1,n} \big] = \sum_{j=1}^{\infty} (1-r)^{j-1}r\sum_{k = j}^{\infty} F_k(n+1|n).
\end{equation}
We can now utilize the relation between the first passage time density and the survival function given in Eq. (\ref{fqrelation-2}) to find
\begin{align}
    \text{Pr}\big[ R\leq T_{n+1,n} \big] &= r \sum_{j=1}^{\infty} (1-r)^{j-1} Q_{j-1}(n+1|n)\\
    &= r\widetilde{Q}_{1-r}(n+1|n).
\end{align}
\\

\noindent
\textbf{Final expression:} 
Putting all the terms together in Eq.~\eqref{SI_meanLn3}, we have
\begin{align}
     \langle L_R(n) \rangle &= \widetilde{Q}_{1-r}(n+1|n) + r\widetilde{Q}_{1-r}(n+1|n)\frac{\widetilde{Q}_{1-r}(n+1|0)}{\widetilde{F}_{1-r}(n+1|0)} \nonumber \\
     &=\widetilde{Q}_{1-r}(n+1|n) + r\widetilde{Q}_{1-r}(n+1|n)\frac{\widetilde{Q}_{1-r}(n+1|0)}{1-r\widetilde{Q}_{1-r}(n+1|0)},
\end{align}
where we have used relation (\ref{fqrelation}). 
Further simplification leads to
\begin{align}
     \langle L_R(n) \rangle &= \widetilde{Q}_{1-r}(n+1|n)\bigg( 1 + r\frac{\widetilde{Q}_{1-r}(n+1|0)}{1-r\widetilde{Q}_{1-r}(n+1|0)}\bigg) \nonumber \\
     &=\frac{\widetilde{Q}_{1-r}(n+1|n)}{1-r\widetilde{Q}_{1-r}(n+1|0)} %= \frac{\widetilde{Q}_{1-r}(n+1|n)}{\widetilde{F}_{1-r}(n+1|0)}
     , \label{SI_main1}
\end{align}
which is Eq. (3) in the main text.

\subsection{Application to random walk}
We now demonstrate Eq. (\ref{SI_main1}) in the case of 1D random walks. As it is evident from Eq. (\ref{SI_main1}), that to compute the mean record age under restart, only the survival or first passage quantities for the restart free processes are required. Many such key quantities are explicitly known in the literature for the random walks. We make use of these existing results to compute the mean record ages under restart. 

\subsubsection{1D unbiased random walk}
First, we consider an unbiased random walk which can hop to the nearest neighbour lattice points with probability $\frac{1}{2}$. For such a random walk, the probability generating function of the discrete first passage time/step is a well known quantity and reads \cite{4}
\begin{equation}
     \widetilde{F}_{z}(n|0) = \biggr( \frac{1-\sqrt{1-z^2}}{z}\biggr)^{n}, \label{unbias1}
\end{equation}
where $n$ is the first passage step. Similarly,
\begin{equation}
     \widetilde{F}_{z}(n+1|n) = \biggr( \frac{1-\sqrt{1-z^2}}{z}\biggr),\label{unbias2}
\end{equation}
One can now directly obtain the the generating function for the survival probability using  Eq.~\eqref{fqrelation}
\begin{equation}
     \widetilde{Q}_{z}(n+1|n) = \biggr( \frac{z-1+\sqrt{1-z^2}}{z-z^2}\biggr).
\end{equation}

We now have all the ingredients to obtain the mean record age $\langle L_R(n) \rangle$. Substituting the expression for the survival probability in Eq. (\ref{SI_main1}), we arrive at the following expression for the mean record age under restart for 1D unbiased random walk
\begin{equation}
     \langle L_R(n) \rangle = \frac{(1-r)^n\bigg( \sqrt{2r-r^2}-r\bigg)}{r(1-\sqrt{2r-r^2})^{n+1}}.
\end{equation}

\subsubsection{1D biased random walk}
We now consider a biased random walk which hops to the right lattice point with probability $p$, and hops to the left with the complementary probability $1-p$. For the case of biased walks, as discussed in the main text, the probability generating function for the first-passage time/step density can be found using standard techniques \cite{4}
\begin{equation}
     \widetilde{F}_{z}(n|0) = \bigg(\frac{1 - \sqrt{1 - 4 p(1 - p)  z^2}}{2(1 - p) z}\bigg)^n,
     \label{bias1}
\end{equation} 
and similarly,
\begin{equation}
     \widetilde{F}_{z}(n+1|n) = \frac{1 - \sqrt{1 - 4 (1 - p) p z^2}}{2(1 - p) z}.
     \label{bias2}
\end{equation} 
It is easy to see that Eqs.~\eqref{bias1}~and~\eqref{bias2} reduce to Eqs.~\eqref{unbias1}~and~\eqref{unbias2} when $p=\frac{1}{2}$. Following the same procedure as before, we use Eq.~\eqref{fqrelation}~and~\eqref{SI_main1} to obtain the following expression for the mean record age under restart 
\begin{equation}
   \langle L_R(n) \rangle = \frac{2^n \bigg(\frac{1 - \sqrt{1 + 4 (-1 + p) p (-1 + r)^2}}{(-1 + p) (-1 + r))}\bigg)^{-n} (1 - 2 p + 
   \sqrt{1 + 4 (-1 + p) p (-1 + r)^2} - 2 r + 2 p r)}{(-1 + \sqrt{
   1 + 4 (-1 + p) p (-1 + r)^2}) r}.
\end{equation}
The above expression was further used to analyze the mean record age as a function of restart probability $r$ [see Fig 2 in the main text].

\section{A universal criterion for restart to expedite record-ages and phase-diagram for RW} \label{S2}
In this section, we derive a criterion which underpins the effect of restart in characterization of the mean record ages. For the 1D symmetric random walk considered above, the mean record age is infinite in the absence of restarts and thus any restart probability $r<1$ would render the mean record age finite. Thus, resetting events always accelerate record breaking events for symmetric random walks. However, this need not be the case in general. An illustrative example is of the biased lattice random walk without restarts, where a ``drift" renders the mean first passage time (and effectively the mean record age) to all the points "downhill" the initial position to be finite. In such cases, it is not apparent why/when restarts should be useful. This is evident also from Fig. 2b, where we have plotted $\langle L_R(4) \rangle$ as a function of restart probability ($r$). The figure shows that while one case restart simply delays the record age, in the other case restart can also shorten the record age. Thus, it is very important to pinpoint the condition which dictates this two-fold behavior.

To see this, a natural way would be to introduce a very small restart probability $r \to 0$ and examine its effects on the mean record age. Expanding Eq.~\eqref{SI_main1} for small $r$ gives 
\begin{align}
   \nonumber \langle L_R(n) \rangle |_{ r \to 0} &=  \frac{\widetilde{Q}_{1-r}(n+1|n)}{1-r\widetilde{Q}_{1-r}(n+1|0)}\Bigg|_{r \to 0}\\
   &\approx \langle L(n) \rangle + \frac{r}{2} \bigg(2 \langle T_{n + 1, n} \rangle \langle T_{n + 1, 0}\rangle + \langle T_{n + 1,n} \rangle - \langle T_{n + 1, n} \rangle^2 - Var(T_{n + 1, n}) \bigg)
   \label{small-r-expansion-mean-L}
\end{align}
where $\langle L(n) \rangle = \langle T_{n+1,n} \rangle$ is the record age for the reset free process and is considered to be finite. Furthermore, we have ignored the terms which are of order $ r^2$ and beyond. The above expression can be written in terms of the coefficient of variation of the random variable $T_{n + 1,n}$, defined as $CV_{n+1,n}= \frac{\text{sd}(T_{n + 1,n})}{\langle T_{n + 1,n}  \rangle}$ [sd stands for `standard deviation']. Following simplifications, we get
\begin{equation}
       \nonumber \langle L_R(n) \rangle |_{r \to 0} 
   \approx \langle L(n) \rangle + r\frac{\langle T_{n + 1,n} \rangle^2}{2} \bigg(2 \frac{\langle T_{n + 1,0}\rangle}{\langle T_{n + 1,n}\rangle} + \frac{1}{\langle T_{n + 1,n} \rangle} -1 - CV_{n+1,n}^2 \bigg).
\end{equation}
In order to have $\langle L_R(n)\rangle < \langle L(n)\rangle$, we must impose
\begin{equation}
    2 \frac{\langle T_{n + 1,0}\rangle}{\langle T_{n + 1,n}\rangle} + \frac{1}{\langle T_{n + 1,n}\rangle} -1 - CV_{n+1,n}^2 <0
\end{equation}
which gives us the general criterion
\begin{equation}
    CV_{n+1,n}^2 > 2 \frac{\langle T_{n + 1,0} \rangle}{\langle T_{n + 1,n}\rangle} + \frac{1}{\langle T_{n + 1,n}\rangle} -1.
\end{equation}
Notably the criterion does not depend on the particular choice of random walk and thus is quite universal. Moreover, it is worth remarking that to understand the effect of restart, it is important to investigate only the statistical metrics for the underlying reset free process and not the reset induced process. 

We can simplify this criterion for the example of 1D biased random walk. As considered in the main text, using translational invariance we have $\langle T_{n + 1,0}\rangle = (n+1)\langle T_{n + 1,n}\rangle$, yielding the criterion
\begin{align}
    \nonumber CV_{n+1,n}^2 &> 2(n+1) + \frac{1}{\langle T_{n + 1,n}\rangle} -1\\
     &> 2n + \bigg\{1 + \frac{1}{\langle T_{n + 1,n}\rangle} \bigg\}. 
     \label{cri-BRW}
\end{align}
Both the mean and coefficient of variation used above are properties of the underlying reset free process and thus depend only on bias $p$ and step $n$. Taking a closer look at Eq. (\ref{cri-BRW}) suggests that the condition is an inequality and then both $p$ and $n$ can not be arbitrary to satisfy the criterion. Thus, in the phase space, spanned by $(p,n)$, Eq. (\ref{cri-BRW}) will naturally give us a seperatrix which can be obtained exactly by solving $CV_{n+1,n}^2 = 2n + \bigg\{1 + \frac{1}{\langle T_{n + 1,n}\rangle} \bigg\}$. See Fig. (2) in the main text where we have plotted this phase-diagram along with the phase-seperatrix. Along this seperatrix, each point will correspond to a \emph{critical} bias value $p^*(n)$, such that for any $p>p^*(n)$, restart can not shorten the age of the $n$-th record. The critical value is then expressed as
\begin{equation}
    p^*(n) = \frac{1}{12} (5 - 2 n + \sqrt{13 + 4 n + 4 n^2}),
    \label{sep}
\end{equation}
which has been used in plotting of the phase diagram in Figure 2a. In the limit of $n\to\infty$, we have
\begin{equation}
     \lim_{n\to\infty} p^*(n) = \frac{1}{12} \Bigg(5 - 2 n \bigg( 1 - \sqrt{ \frac{13}{4n^2} + \frac{1}{n} + 1}\bigg)\Bigg) =  \frac{1}{12} \Bigg(5 - 2 n \bigg( 1 - \sqrt{  1 + \frac{1}{n}}\bigg)\Bigg) ,
    \label{sep1}
\end{equation}
where, using the approximation $\sqrt{  1 + \frac{1}{n}} \approx 1 + \frac{1}{2n}$, we get  
\begin{equation}
    \lim_{n\to\infty} p^*(n) = \frac{1}{12} \Bigg(5 + 2 n \bigg( \frac{1}{2n}\bigg)\Bigg).
    \label{sep2}
\end{equation}
This allows us to see that 
\begin{equation}
    \lim_{n\to\infty} p^*(n) = \frac{1}{2}.
    \label{sep3}
\end{equation}
Equation~\eqref{sep3} establishes that geometric restart loses the ability to shorten record ages for biased random walks, as the record number becomes larger. 

\begin{figure}[h]
    \centering
    \includegraphics[width=0.6\columnwidth]{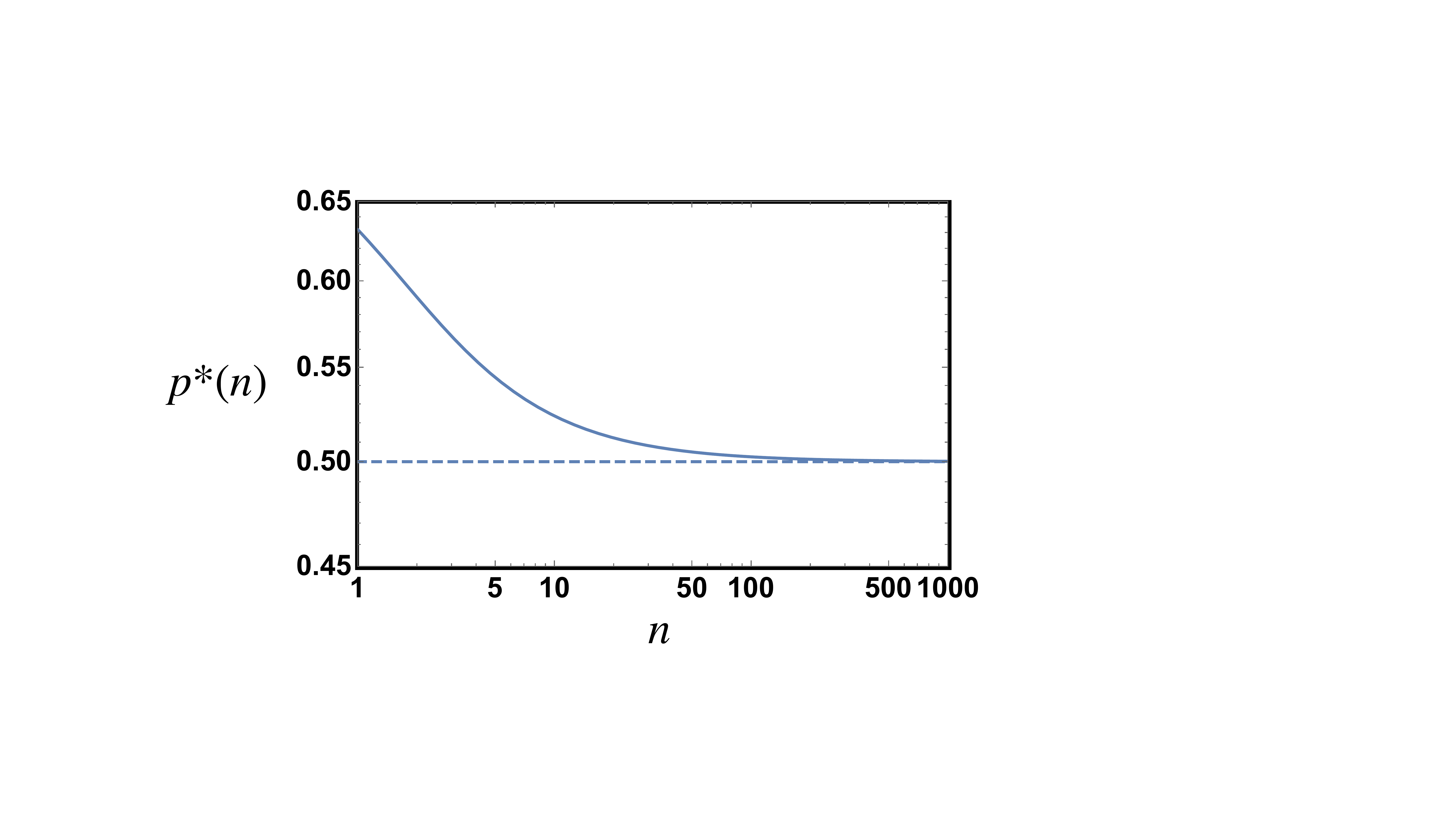}
    \caption{A plot of $p^*(n)$, as expressed in Eq.~\eqref{sep}, as a function of record index $n$. For large $n$, restart fails to shorten record ages in biased walks, and $\lim_{n\to\infty} p^*(n) = \frac{1}{2}$, corroborating the asymptotic relation in Eq.~\eqref{sep3}.}
    \label{fig:SI2}
\end{figure}

\textcolor{black}{\section{Effect of restart on Record ages when $X(t)$ is strictly non-negative}}
\noindent
\textcolor{black}{In certain applicable scenarios, the time-series of interest $X(t)$ may not take negative values and may strictly remain non-negative at all times. For example, let $X(t)$ denote the mass of an aggregate at a nucleation site, where the arrival of unit masses (monomers) follows a geometric distribution (Markovian) with arrival probability $p$ at each time-step. This is an important example of a scenario where the time-series of interest remains non-negative. Furthermore, consider that there is a clearance mechanism in play which, independently with probability $r$, acts on the system and resets the mass at the site to $0$ (akin to restart). }

\textcolor{black}{Assume the mass at the site is $0$ to begin with, and let us look at the age of the $n$th record, i.e. the time taken for the mass at the nucleation site to reach $n+1$ for the first time, after it has reached a mass $n$. Since this follows a geometric distribution, the mean record age in the absence of restart is simply $1/p$. Similarly, the coefficient of variation ($CV$) is $(1-p)$. Plugging them into Eq.~(6) from the main text, we get
\begin{equation}
    1-p>2n+1+p
\end{equation}
where it is clear that the inequality can never be satisfied, and thus, restart cannot expedite the creation of records for any $n>0$ and any value of $p$ in $[0,1]$.
}

\begin{figure*}
    \centering
    \includegraphics[width=\columnwidth]{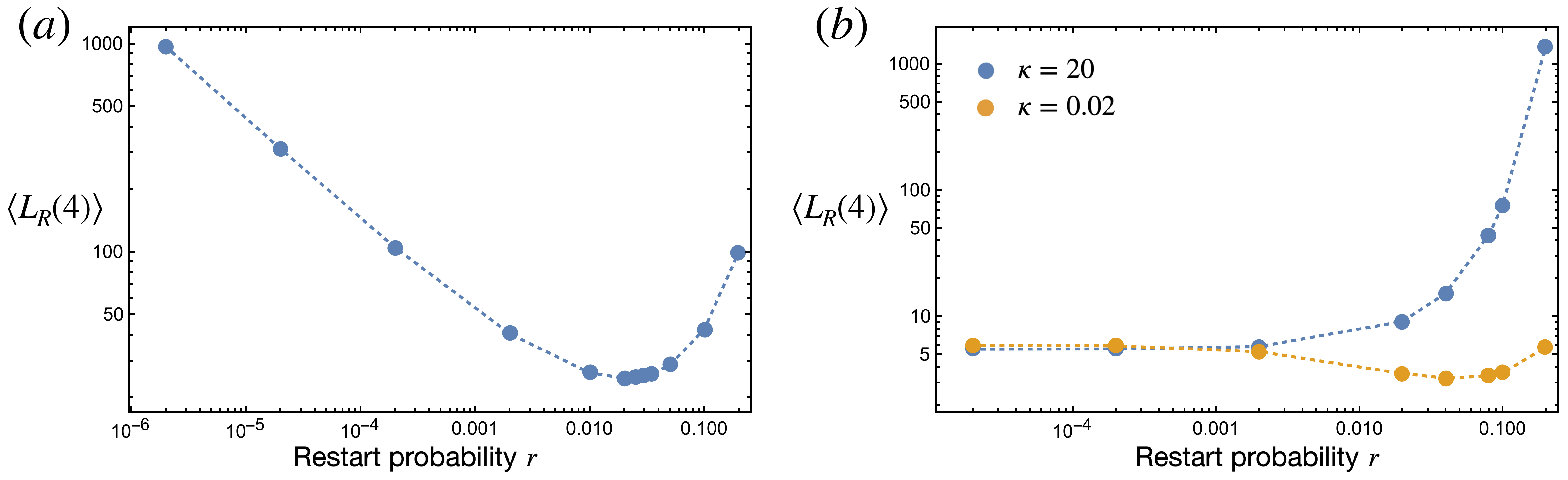}
    \caption{\textcolor{black}{Mean age of the $4$th record as a function of restart probability (restart steps are taken from geometric distribution with parameter $r$), when the arrival times for the masses are drawn from (a) the Zipf distribution $P(k) = \frac{k^{-a}}{\zeta(a)}$, where $\zeta(a)$ is the Riemann-Zeta function and the value of $a$ was chosen to be $3/2$. Despite the underlying process having a diverging mean record age, introduction of restart makes it finite. (b) In this case, arrival times are drawn from the discretized Gamma distributions $\lceil \Gamma(x) \rceil$, where $\Gamma(x)$ is the Gamma distribution with shape parameter $\kappa$, and scale parameter given by $\frac{5}{\kappa}$. For $\kappa = 0.02$ (high CV). It is quite evident that restart is able to shorten record ages, whereas for the case of low CV in $\kappa = 20$, restart only leads to longer record ages. See the text for further details.} }
    \label{fig:record_age_nonnegative}
\end{figure*}

\textcolor{black}{On the other hand, suppose the distribution of times for a unit mass to arrive is not a geometric distribution, and is given by a heavy-tailed distribution with a diverging mean. Clearly, the mean record ages in the absence of any clearance mechanism will diverge. However, in the presence of clearance mechanisms, where a new arrival time is drawn after each clearance (i.e. restart) event, all the mean record ages become finite. The intuition behind this is that the contribution due to rare but very long arrival times will be impeded by restart, thus reducing/shortening the record age. This intuition is verified in simulation results presented in Fig.~\ref{fig:record_age_nonnegative}(a), where the mean age of the $4$th record (mean time taken for an aggregate of size $5$ to be formed after the formation of an aggregate of size $4$ for the first time) as a function of restart probability (restart steps are taken from geometric distribution), when the arrival times are drawn from the Zipf distribution $P(k) = \frac{k^{-a}}{\zeta(a)}$, where $\zeta(a)$ is the Riemann-Zeta function and the value of $a$ was chosen to be $3/2$. Despite the mean of this distribution diverging, it is evident from Fig. \ref{fig:record_age_nonnegative}(a) that the  shattering/restart events render the mean record age finite.}

\textcolor{black}{Generically, for arrival time distributions which have finite moments, the criterion derived in Eq.~(6) from the main text plays an important role in gauging whether record ages can be shortened by such restart events. In Fig.~\ref{fig:record_age_nonnegative}(b), we present another result for the mean age of the fourth record under restart, where the arrival times are drawn from two different discretized Gamma distributions $\lceil \Gamma(x) \rceil$, where $\Gamma(x)$ is the Gamma distribution with shape parameter $\kappa$, and scale parameter given by $\frac{5}{\kappa}$, and $\lceil \cdot \rceil$ denoting the ceiling function. We choose $\kappa=20$ and $\kappa=0.02$ for the two distributions respectively. While the mean is roughly the same for both the discretized Gamma distributions, the coefficient of variation is different (high for $\kappa=0.02$ and much lower for $\kappa=20$). As the criterion in Eq.(6) is satisfied for the distribution with $\kappa=0.02$ for $n=4$, we see indeed that the record age can be reduced via restart. On the other hand, in the case where the arrival time is given by the discretized Gamma distribution with $\kappa=20$, the criterion is not satisfied and restart evidently prolongs the record age.}

\textcolor{black}{In summary, the validity of the criterion in Eq. (6) goes beyond the processes where the observables can only take non-negative values, and allows us to quantitatively determine whether restart can expedite record creation in a fairly general setting.}\\

\section{Record ages under Arbitrary restart time density -- Beyond Markovian resetting} \label{S3}
In this section, we sketch out the steps for computing record ages for stochastic processes under arbitrary restart time which may not be Markovian (geometric). To this end, we recall the renewal equation discussed in the main text for the age of records under generic restart time density [see Eq. (7) and Fig. (3) in the main text]
\begin{equation}
    L_R(n) = \begin{dcases}
    L(n),& \text{if } L(n) < R_{\delta_n}\\
    R_{\delta_n} +  T^{R}_{n+1,0}   & \text{otherwise}
    \end{dcases}
\label{SI_arb_reset}
\end{equation} 
where $R_{\delta_n} = \{R - \delta_n | R>\delta_n\}$ is the forward renewal/recurrence time -- often called residue time and $\delta_n$ (backward renewal/recurrence time -- often called aging time) is given by \begin{equation}
    \delta_n = \{ T_{n,0} |T_{n,0} <R\},
\end{equation}
which is distributed according to
\begin{align}
    P_{\delta_n}(k)%\text{Pr}(\delta_X=k)=
%&=F_k(n|0)\frac{\sum_{m=k+1}^{\infty}\text{Pr}\left(R=m\right)}{\text{Pr}\left(T_{n,0} < R\right)}, \\
&=F_k(n|0)\frac{\sum_{m=k+1}^{\infty}P_R(m)}{\text{Pr}\left(T_{n,0} < R\right)},\label{SI_dx}
\end{align}
where $P_R(m)$ is the restart time density (but not necessarily geometric). It is important to note that $R_{\delta_n}$, for geometrically distributed resets, has the same distribution as $R$, and thus one simply recovers Eq.\eqref{SI_renewal}. For the generic case, we can write the following, by definition,
\begin{equation}
    P_{R_{\delta_n}}(k) = \text{Pr}(R-\delta_n = k|R-\delta_n>0),
\end{equation}
which can be explored further using the definition of conditional probability,
\begin{equation}
    P_{R_{\delta_n}}(k) = \frac{\text{Pr}(R-\delta_n = k~ \&~ R-\delta_n>0)}{\text{Pr}(R-\delta_n>0)}.
\end{equation}
If we consider positive values of $k$, then the condition $R-\delta_n>0$ is redundant, and we have
\begin{equation}
    P_{R_{\delta_n}}(k) = \frac{\text{Pr}(R-\delta_n = k)}{\text{Pr}(R-\delta_n>0)}
\end{equation}
which can be written as
\begin{equation}
    P_{R_{\delta_n}}(k) = \frac{\sum_{m}\text{Pr}(R = k + m ~ \& ~ \delta_n = m)}{\text{Pr}(R-\delta_n>0)}.
    \label{Rn-1}
\end{equation}
Using the distribution of $\delta_n$ [see Eq. (\ref{SI_dx})], the joint density in the numerator of the above equations $\text{Pr}(R = k + m ~ \& ~ \delta_n = m)$ can be expressed as
\begin{equation}
  \text{Pr}(R = k + m ~ \& ~ \delta_n = m) =   F_m(n|0)\frac{\text{Pr}\left(R=k + m \right)}{\text{Pr}\left(T_{n,0} < R \right)}~\underbrace{\text{Pr}[R>m]}_{=1}.
\end{equation}
Summing over all admissible values of $m$, we have
\begin{equation}
  \sum_m \text{Pr}(R = k + m ~ \& ~ \delta_n = m) =    \frac{\text{Pr}\left(R = k + T_{n,0} \right)}{\text{Pr}\left(T_{n,0} < R \right)}
\end{equation}
which allows us to write the distribution of $R_{\delta_n}$ as [following Eq. (\ref{Rn-1})]
\begin{equation}
    P_{R_{\delta_n}}(k) = \frac{\text{Pr}\left(R -  T_{n,0} = k  \right)}{\text{Pr}\left(R - T_{n,0} > 0 \right) \cdot \text{Pr}(R-\delta_n>0)}.
\end{equation}
Thus, we have obtained the exact distribution for $R_{\delta_n}$ which essentially give us all the ingredients to obtain the record ages under arbitrary restart using Eq. (\ref{SI_arb_reset}). For example, to obtain the mean record age, we take expectation on both sides of Eq. (\ref{SI_arb_reset}) to obtain
\begin{equation}
    \langle L_R(n) \rangle = \langle \text{min}\{L(n),R_{\delta_n}\}  \rangle + \text{Pr}\big(R_{\delta_n}\leq L(n)\big) \langle T^{R}_{n+1,0} \rangle,
    \label{arb_meanLn1SI}
\end{equation}
which is Eq.~(10) in the main text. Clearly, the terms on the RHS can be computed from the statistical distributions of the underlying process (such as $F_k(i|j)$ for $T$ and $ P_{R_{\delta_n}}(k)$ for $R_{\delta_n}$). \\

\noindent
For completeness, we write below the exact relations [which are used to compute the first and second term on the RHS in Eq. (\ref{arb_meanLn1SI})] for two arbitrary random variables $z$ and $y$ \cite{2,3}. We start by noting that $\text{Pr}\left(\text{min}\left(z,y\right)>\ell\right)=\text{Pr}\left(z>\ell\right)\text{Pr}\left(y>\ell\right)$ and thus
\begin{align}
\left\langle \text{min}\left(z,y\right)\right\rangle&=\sum_{\ell=0}^{\infty}\text{Pr}\left(\text{min}\left(z,y\right)>\ell\right) \nonumber \\
&= \sum_{\ell=0}^{\infty} \left(\sum_{k=\ell+1}^{\infty}P_z\left(k\right)\right) \left(\sum_{m=\ell+1}^{\infty}P_y\left(m\right)\right) \nonumber \\
&= \sum_{\ell=0}^{\infty} Q_\ell^{(z)} Q_\ell^{(y)}
, \label{Mean min formula}
\end{align}
where note that $Q_\ell^{(z)}$ and $Q_\ell^{(y)}$ are the survival functions for the random variables $z$ and $y$ respectively so that
\begin{align}
    Q_\ell^{(z)} &=\text{Pr}(z>\ell), \label{Survival N} \\
    Q_\ell^{(y)}&=\text{Pr}(y>\ell) \label{Survival R}.
\end{align}
The conditional probability [second term on the RHS in Eq. (\ref{arb_meanLn1SI})] can also be computed easily 
\begin{align}
\text{Pr}\left(z < y\right)=\sum_{\ell=0}^{\infty}P_{z}(\ell)\sum_{m=\ell+1}^{\infty}P_{y}(m), \label{denominator}
\end{align}
and $\langle T^{R}_{n+1,0} \rangle$ is the mean first passage step under restart given in Eq. (\ref{FPUR-dis}).

\section{Simulation details of Aggregation-shattering processes} \label{S4}
\begin{figure}
    \centering
    \includegraphics[width=0.95\columnwidth]{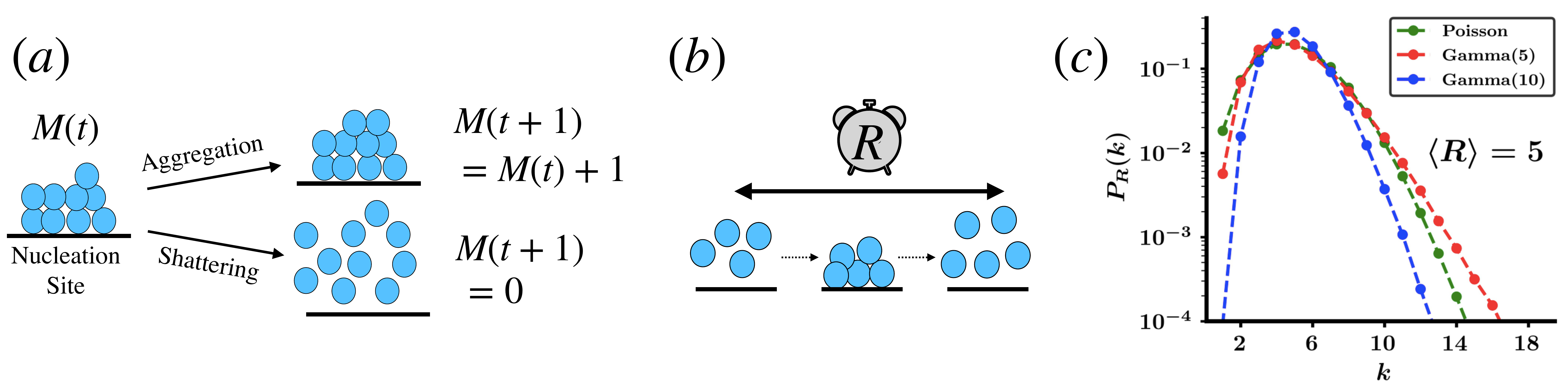}
    \caption{Application of the formalism to aggregation-shattering processes. (a) A schematic of the aggregation-shattering process at a nucleation site. In each time-step, with some probability a monomer can arrive and the nucleation site and form a larger aggregate. However, each time-step also carries the possibility of shattering, reminiscent of clearance events in biological systems, which render the mass at the nucleation site to go to $0$. (b) A schematic depicting the random variable $R$, which denotes the time between two consecutive shattering events. (c) A plot of the three non-geometric distributions of $R$ (described in Sec.~\ref{S4}), which were used to produce Fig.~4(b) in the main text.}
    \label{fig:SI_1}
\end{figure}

To illustrate the power of our universal approach, in the main text, we have discussed an important application 
where some of these general results can be directly applied. We look into a stochastic mass transport model namely an aggregation-shattering process \cite{5,6,7,8}. To further illustrate, let us consider a model of aggregate formation at a given nucleation site, where we keep track of the aggregated mass $M(t)$ at that site, as a function of time. Masses aggregate when monomers arrive at the nucleation site with certain probability followed by shattering events which reset the mass index at the nucleation site. In the example considered, the inter-arrival times between two monomers follow a geometric distribution. Upon the arrival of a monomer on the nucleation site, the monomer sticks to the cluster of masses at the nucleation site (an aggregation event), leading to an increase in the mass at the site by unit one. However, `clearance' occurs at random times, which might be drawn from arbitrary non-geometric distributions $P_R(k)$. Clearance leads to the shattering of the cluster at the nucleation site, rendering $M(t)=0$ in that time-step [see Fig. (\ref{fig:SI_1})]. 

We are interested in computing the record age statistics $ L_R(n)$ of the time-series $M(t)$. In the context of aggregation-shattering processes, $L_R(n)$ denotes the random time taken for a cluster of mass $n+1$ to be formed after the formation of an aggregate of mass $n$ for the first time. Evidently, $L_R(n)$ sheds light on the rate of growth of a cluster of size $n$.

For the preparation of Fig.~$(4)$ in the main text, the following representative shattering time distributions were used:
\begin{itemize}
    \item Poisson distribution: $P_R(k) = \frac{\lambda^k e^{-\lambda}}{k!}$, which has a mean of $\lambda$.
    \item Discretized Gamma distribution: $ \lceil \Gamma(x) \rceil$, where $\Gamma(x)$ is the Gamma distribution with shape parameter $\kappa$, and scale parameter given by $\frac{\langle R \rangle}{\kappa}$. In Fig.~$(4)$, we use $\kappa=5$ and $10$ for the red and blue curves respectively.
\end{itemize}

In Fig.~4(b) of the main text, we plotted the mean age of the fourth record ($\langle L_R(4) \rangle$) in the aggregation-shattering process, i.e., the mean time taken for an aggregate of size $5$ to be created, after the aggregate size reaches $4$ for the first time, using the above three different restart distributions (also plotted in Fig.~\ref{fig:SI2}(c)). The mean record ages were plotted as a function of different mean restart times, using Eq.~\eqref{arb_meanLn1SI} (or equivalently, Eq.~10 from the main text).

\end{document}